\documentclass[prl,superscriptaddress,amsmath,amssymb,floatfix,twocolumn]{revtex4-1}
\usepackage{graphicx,color}
\usepackage{amsmath,amsfonts,amssymb}
\usepackage{float}
\usepackage[colorlinks,linkcolor=blue,anchorcolor=blue,urlcolor=blue,citecolor=blue]{hyperref}

\begin{document}

\title{Anomalous topological edge states in non-Hermitian piezophononic media}

\author{Penglin Gao}
\email[Corresponding author.\\]{pgao@pa.uc3m.es}
\affiliation{Department of Physics, Universidad Carlos III de Madrid, ES-28916 Legan\`es, Madrid, Spain}
\author{Morten Willatzen}
\affiliation{Department of Photonics Engineering, Technical University of Denmark, Kongens Lyngby, DK‐2800 Denmark}
\affiliation{Beijing Institute of Nanoenergy and Nanosystems, Chinese Academy of Sciences, No. 30 Xueyuan Road, Haidian District,Beijing 100083, China}
\author{Johan Christensen}
\email[Corresponding author.\\]{johan.christensen@uc3m.es}
\affiliation{Department of Physics, Universidad Carlos III de Madrid, ES-28916 Legan\`es, Madrid, Spain}
\date{\today}

\begin{abstract}
The bulk-boundary or bulk-edge correspondence is a principle relating surface confined states to the topological classification of the bulk. By combining non-Hermitian ingredients in terms of gain or loss with media that violate reciprocity, an unconventional non-Bloch bulk-boundary correspondence leads to unusual localization of bulk states at boundaries$-$a phenomenon coined non-Hermitian skin effect. Here we \textcolor{black}{numerically} employ the acoustoelectric effect in electrically biased and layered piezophononic media as a solid framework for non-Hermitian and nonreciprocal topological mechanics in the MHz regime. Thanks to a non-Hermitian skin effect for mechanical vibrations, we find that the bulk bands of finite systems are highly sensitive to the type of crystal termination, which indicates a failure of using traditional Bloch bands to predict the wave characteristics. More surprisingly, when reversing the electrical bias, we unveil how topological edge and bulk vibrations can be harnessed either at the same or opposite interface. Yet, while bulk states are found to display this unconventional skin effect, we further discuss how in-gap edge states in the same instant, counterintuitively are able to delocalize along the entire layered medium. We foresee that our predictions will stimulate new avenues in echo-less ultrasonics based on exotic wave physics.
\end{abstract}

\maketitle

The discovery of topological phases has spurred tremendous efforts in photonics \cite{Ozawa:RevModPhys2019} and phononics \cite{Zhang:CommPhys2018} by pushing forward investigations into defect-immune classical systems. The bulk-edge correspondence (BEC) plays a crucial role since it establishes a bridge between topological edge states and the Bloch band topology \cite{Ezawa:PRB2013, Benalcazar:Science2017}. \textcolor{black}{Thanks to this principle, we have witnessed many viable routes to engineer topologically protected waves control in recent years \cite{Haldane:PRL2008,Miniaci:PRX2018,cornerNJU,ZZhang,JK}}. Lately, at an equally active frontier, do we find parity-time synthetic materials, that enable new wave characteristics by cleverly crafted loss and gain components. The Hamiltonians of such non-Hermitian systems permit exceptional singularities, where the eigenvalues together with their corresponding eigenstates simultaneously coalesce, leading to highly unusual functionalities such as enhanced sensing and one-way invisibility cloaks \cite{Hodaei:Nature2017}.\\
Meanwhile, the combination of non-Hermitian and topological physics is attracting considerable interest \cite{Lee:PRL2016, Gong:PRX2018, Kunst:PRL2018, Longhi:PRL2019, ZhangZW:PRL2019, Zhao:Science2019, Xue:PRL2020, Brandenbourger:NC2019, Ghatak:arXiv2019, Zhu:PRResearch2020}, partially stemming from the controllable non-Hermitian elements, capable to provide additional freedom to harness topological phases \cite{ZhangZW:PRL2019, Zhao:Science2019}. But most importantly, such systems open yet unexplored avenues in physics that strongly challenge the BEC by breaking with its most common conventions. The so-called non-Hermitian skin effect \cite{Yao:PRL2018, Song:PRL2019, Okuma:PRL2020, Xiao:NatPhys2020, Helbig:NatPHys2020, Weidemann:Science2020}, a phenomenon featured by abnormal bulk states localization, has thoroughly undermined our understanding of the Bloch band topology. Simply put, based on nonreciprocal hopping of electrons that move in preferential directions, translational symmetry breaking is brought forward and consequently the failure of conventional Bloch bands stands before. Beyond the rich physics involved, extensions of the skin effect to classical systems may provide new opportunities in reflectionless guiding of waves based on the combination of nonreciprocal non-Hermitian media and topological ingredients. \\
\textcolor{black}{In this Letter, we propose a continuum mechanical approach to access non-Hermitian topology. We numerically employ piezoelectric semiconductors, which when electrically biased generate a nonreciprocal response owing to the acoustoelectric effect.} Thanks to the intrinsic electron-phonon interaction, sound amplification or attenuation is taking place in response to an appropriately applied electric field, which is interesting for various appealing applications \cite{White:JAP1962,Christensen:PRL2016, Merkel:PRApplied2018}. Thus, this accessible piezophononic platform is employed to mimicking the non-Hermitian Su-Schrieffer-Heeger (SSH) model \cite{Su:PRL1979, Yao:PRL2018} for mechanical vibrations. We predict a ultrasonic counterpart to the non-Hermitian skin effect and the ensuing failure of the Bloch band topology. By reversing the applied field and tuning the attenuation, we show the ability to engineer the confinement of bulk and edge states as one desires. To be exact, we find that topological edge states can even become fully delocalized along the entire crystal bulk, which together with the aforementioned skin states opens horizons of possibilities in non-Hermitian waves engineering.\\
\begin{figure}[htbp]
	\centering
	\includegraphics[scale=1.0]{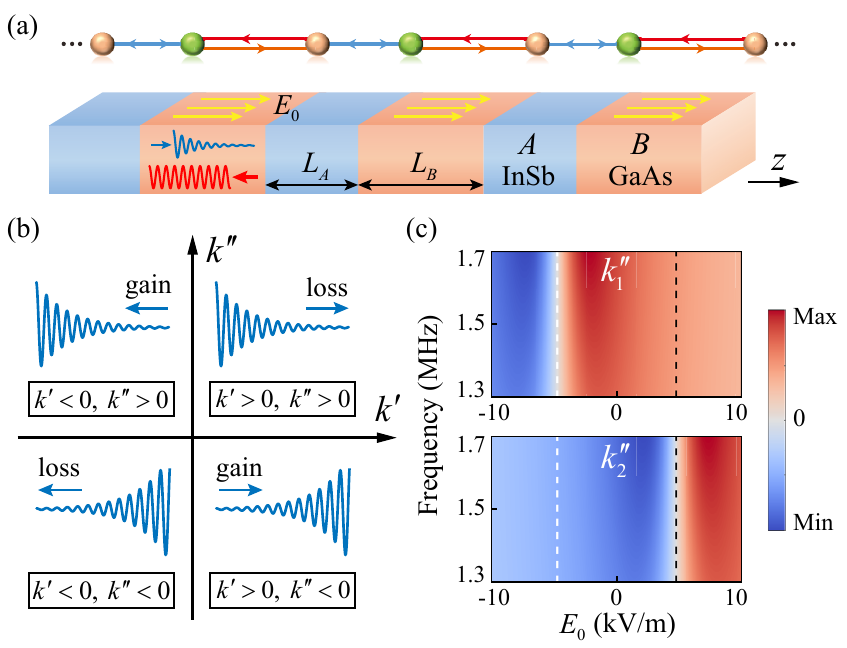}
	\caption{Mechanical SSH model with nonreciprocal coupling. (a) A piezophononic non-Hermitian system, electrically biased with electric field $E_{0}$. The atomic chain illustrates how the continuum system maps to a non-Hermitian SSH model. (b) Modal space, showing the physical meaning of the complex modal solutions $k=k'+ik''$. (c) The imaginary parts of the two modal solutions, $k''_{1}$ and $k''_{2}$, are plotted as a function of $E_{0}$ for frequencies of interest. The dashed lines indicate the Cherenkov threshold where either $k''_{1}=0$ or $k''_{2}=0$, indicating perfect one-way transmission.}
	\label{fig1}
\end{figure}
We begin with a one-dimensional (1D) piezophononic layered system as shown in Fig. \ref{fig1}(a), whose unit cell contains two different materials of length $L_{A}$ and $L_{B}$, where the subscripts $A$ and $B$ denote InSb and GaAs, respectively. \textcolor{black}{The two crystals of the wurtzite family are stacked alongside their hexagonal symmetry axis ($z$-direction), along which, via subwavelength electrodes, a dc electric field $E_{0}$ is applied to the GaAs layers only [see yellow arrows in Fig. \ref{fig1}(a)], to constitute a non-Hermitian activation of the piezophononic medium}. In a small signal approximation the dispersion of longitudinal vibrations is expressed as \cite{White:JAP1962}
\begin{equation}
\label{dispersion_relation}
\rho \omega^2=k^2 \left\{ c + \frac{e^2}{\varepsilon}  \frac{1-\gamma\frac{k}{k_0} +i \frac{\omega}{\omega_d} (\frac{k}{k_0})^2} {1-\gamma \frac{k}{k_0} + i \left [\frac{\omega_c}{\omega} + \frac{\omega}{\omega_d} (\frac{k}{k_0})^2 \right] } \right\},
\end{equation}
where the involved quantities are the frequency $f=\omega/2\pi$, the wave number $k_0=\omega/v_0$, the stiffness $c$, the mass density $\rho$, the piezoelectric constant $e$, and the electric permittivity $\varepsilon$. Moreover, to simplify the equation we have introduced the dielectric relaxation frequency $\omega_c=\sigma/\varepsilon$ with conductivity $\sigma$, the diffusion frequency $\omega_{d}=v_{0}^{2}/d_{n}$ with electron diffusion constant $d_{n}$ and sound velocity $v_{0}=\sqrt{c/\rho}$, and the drift parameter $\gamma=-\mu_{n}E_{0}/v_{0}$ with electron mobility $\mu_{n}$. When the electric field is turned on, nonreciprocity kicks in comprising loss along one path with gain in the opposing direction as rendered in Fig. \ref{fig1}(b). By substituting $k/k_0\approx\pm1$ into Eq. (\ref{dispersion_relation}), here $k=k'+ik''$, we obtain two modal solutions denoted as $k_{1}$ and $k_{2}$ representing the forward ($k'_{1}>0$) and backward ($k'_{2}<0$) propagating modes, respectively. As an example, we perform a calculation with specific parameters of GaAs: $\rho=5320\,\text{kg} / \text{m}^{3}$, $c=85.5\,\text{GPa}$, $\varepsilon=10.89$, $e=-0.16\,\text{C} / \text{m}^{2}$, and $\mu_n=0.85\,\text{m}^2 / (\text{V} \cdot \text{s})$. In Fig. \ref{fig1}(c), we plot their imaginary parts $k''_{1}$ and $k''_{2}$ as a function of $E_{0}$, which display a remarkable nonreciprocal response within the frequency range studied. Of particular interest is the Cherenkov threshold $\gamma=\pm 1$ [dashed lines in Fig. \ref{fig1}(c)], where the drift velocity equals the sound speed, leading to perfect nonreciprocity, i.e., lossy wave propagation in one direction and unattenuated behaviour in the opposite one. To be specific, when $E_{0}=-v_{0}/\mu_{n}$, forwardly directed sound propagation is entirely attenuation free ($k''_{1}=0$, see white dashed lines) while the opposing mode displays some losses as rendered in the $k''_{2}$ plot. If we are to swap the sign of the electric field, i.e. $E_{0}=v_{0}/\mu_{n}$, the scenario is reversed as indicated by the black dashed lines. \\
In order to compute the complex scattering properties of the layered piezophononic medium illustrated in Fig. \ref{fig1}(a), we employ a directional dependent transfer matrix method for both periodic and finite configurations comprising both free and rigid boundary conditions. This method consists in transferring the normal displacement $u$ and stress $\tau$ from the right boundary to the left one, i.e.,
\begin{equation}
\label{T-matrix}
\begin{bmatrix} u_{r} \\ \tau_{r} \end{bmatrix}
=\textbf{T} \begin{bmatrix} u_{l} \\ \tau_{l} \end{bmatrix},
\end{equation}
where $\textbf{T}$ denote the transfer matrix and the subscripts ($l$) and ($r$) indicate their respective boundaries, left and right. For periodic systems, e.g., considering the displacement only, the translational symmetry requires $u_{r}=e^{ik_{B}L} u_{l}$, where $k_{B}$ and $L$ denote the Bloch wave number and the lattice period, respectively, which leads to a secular equation, $\det(\textbf{T}-e^{ik_{B}L}\textbf{I})=0$, to solve an eigenvalue problem comprising Bloch waves. Likewise we can construct a finite system containing $N$ unit cells (for details, [\onlinecite{suppl}]).  
\begin{figure*}
	\centering
	\includegraphics[scale=0.90]{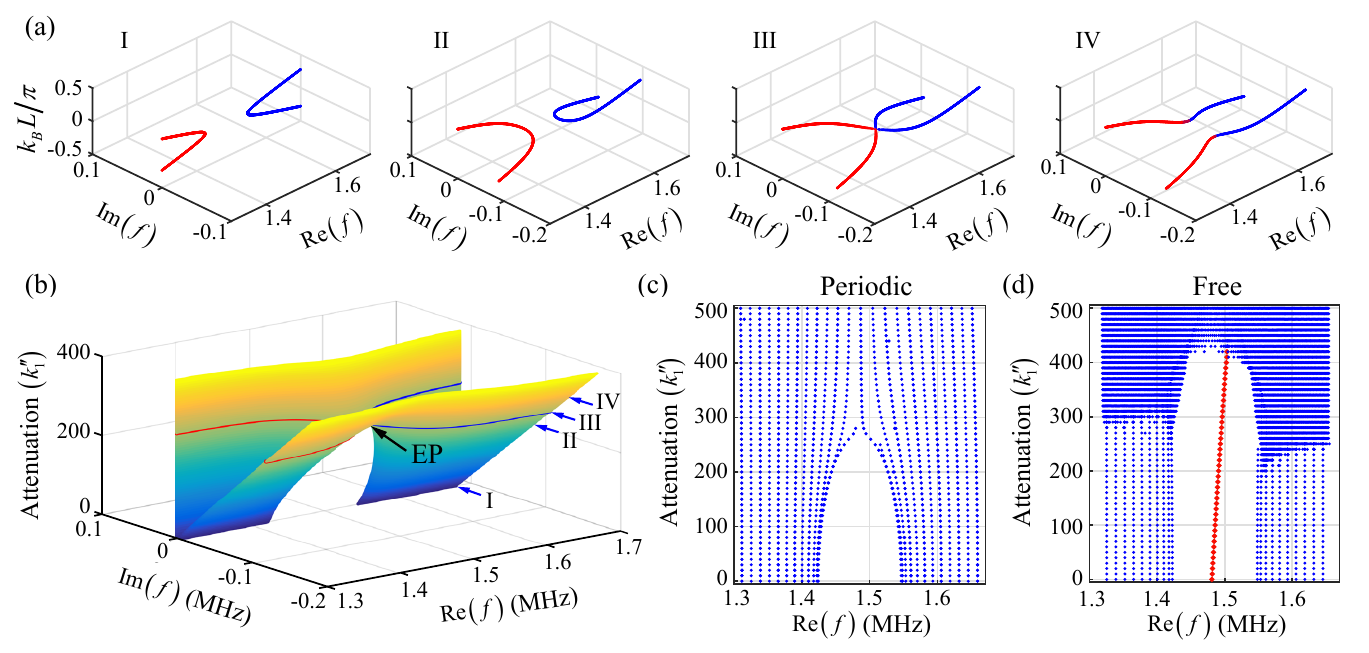}
	\caption{Breakdown of the conventional bulk-edge correspondence. (a) A group of non-Hermitian band diagrams are plotted in the complex frequency plane as a function of the Bloch wave number $k_{B}$. As indicated in panel (b), the selected phases I, II, III and VI are marked at specific attenuation levels $k''_{1}=$ 0, 200, 266, and 300 $m^{-1}$, respectively. Computed eigenfrequency spectra for a finite array (containing $N=50$ unit cells) when subject to periodic (c) and free (d) boundary conditions. For clarity, the in-gap edge states in (d) are highlighted by red dots.}
	\label{fig2}
\end{figure*} 
As rendered in  Fig. \ref{fig2} we will consider both scenarios to unveil the breakdown of the conventional BEC in piezophononic layered media. As computations display in Fig. \ref{fig2}(a) and Fig. \ref{fig2}(b), we firstly compute a group of Bloch bands with specific parameters: $L_{A}=0.5\,\text{mm}$, $L_{B}=2.0\,\text{mm}$, and $\rho=5770\,\text{kg}/\text{m}^3$, $c=47\,\text{GPa}$ for InSb. In the Hermitian limit with zero attenuation (phase I), a purely real band gap is opened due to the impedance contrast in the periodic crystal. \textcolor{black}{When increasing the one-way attenuation $k''_{1}$ [see Fig. \ref{fig2}(b)], imaginary eigenfrequencies set in while the real valleys coalesce towards an exceptional point (EP), phase III. Fig. \ref{fig2}(b) clearly shows this non-Hermitian phase transition whose EP resides at $k''_{1}=266\,\text{m}^{-1}$. This particular phase displays the most unusual topological feature in finite systems as we will discuss later [\onlinecite{suppl}], in general however, while the externally applied electric field must satisfy the Cherenkov threshold, appropriate electrical or optical carrier density manipulation ensures the reaching of the necessary attenuation levels \cite{Merkel:PRApplied2018}}. Indeed, Fig. \ref{fig2}(a) displays with clarity how the added unidirectional loss serves as the control parameter to close and reopen the complex bands, across phases II to IV. Next, we consider a finite crystal of length $N=50$. First, we apply free boundary conditions (FBCs) and in the second case, we consider periodic boundary conditions (PBCs), hence, the latter is thus regarded as a superlattice. The eigenfrequencies are computed against the attenuation strength $k''_{1}$ where it immediately stands out that the crystal with free interfaces host an edge state, \textcolor{black}{which due to multiple scattering is offset the exact mid-gap frequency}, as rendered by red dots in Fig. \ref{fig2}(d). As expected, the periodic system considered [Fig. \ref{fig2}(c)] displays the mode coalescence at phase III equivalent to the former discussion. By interchanging the boundaries to free instead, Fig. \ref{fig2}(d) displays how the phase transition now is pushed back at about $k''_{1}=430\,\text{m}^{-1}$, far from the predicted value mentioned earlier. This anomalous feature is just one among many unusual non-Hermitian characteristics that we are going to discuss in what follows.\\
To shed more light on non-Hermitian topology, we now intend to introduce another factor, the impedance ratio $Z_{A}/Z_{B}$, which is at least equally important in manipulating topological phases when compared to the attenuation. 
\begin{figure}[htbp]
	\centering
	\includegraphics[scale=1.0]{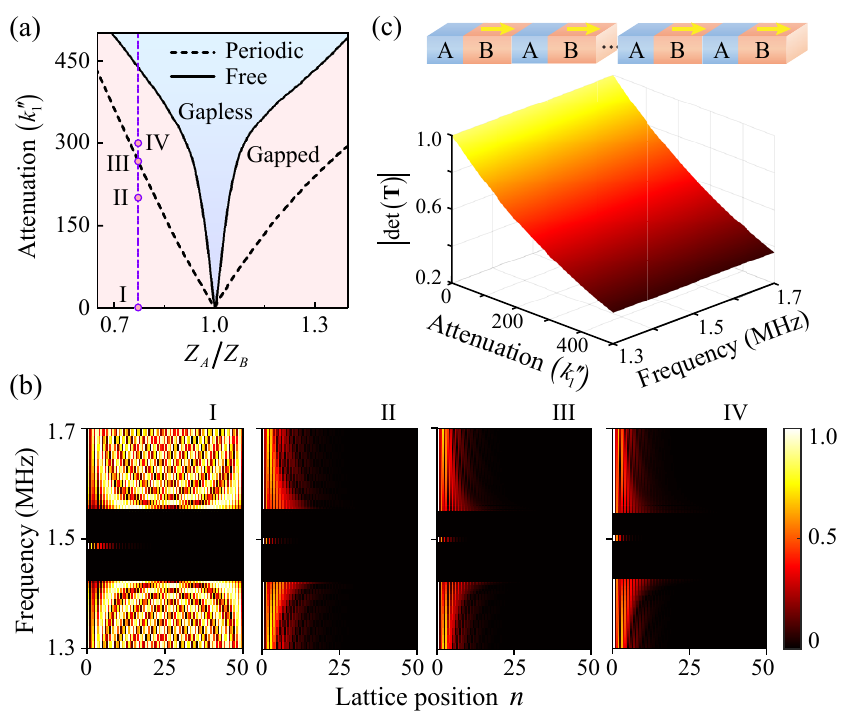}
	\caption{(a) Phase diagram comprising two variables, the impedance ratio $Z_{A}/Z_{B}$ and attenuation $k''_{1}$ (in units $\text{m}^{-1}$) in a forward-biased crystal. The dashed and solid lines mark the phase transition points calculated under periodic and free boundary conditions, respectively. Background colors are used to differentiate the gapped (pink) from the gapless (blue) phase when $N=50$. The purple vertical dashed line corresponds to the phases highlighted in Fig. \ref{fig2} for the piezophonic medium made of InSb and GaAs (constant impedance). (b) Numerically calculated eigenfields for the selected phases as indicated in panel (a). (c) A surface plot mapping the magnitude of the transfer matrix determinant $\left| \det{(\textbf{T})} \right|$.}
	\label{fig3}
\end{figure}
In order to realize such impedance tuning, one is able to replace the InSb constituent of the crystal with a different material, or to structure it for a desired effective impedance $Z_{A}$, in the vein of metamaterials. Based on the same numerical approach discussed before, we compute a phase diagram as a function of $k''_{1}$ and $Z_{A}/Z_{B}$ as shown in Fig. \ref{fig3}(a) for a forward-biased (positive electric field) piezophononic medium. The dashed line indicates the phase boundaries as predicted by the complex Bloch band computations, while the  solid one is obtained from the eigenfrequency spectra of finite systems. Both degeneracies merge at the lossless (Hermitian) limit when the impedance contrast vanishes $Z_{A}/Z_{B}=1$. According to the Bloch band topology, the conventional BEC predicts the formation of topological edge states up until reaching the phase transition point III. However, according to our finite-size computations, beyond this point but below the solid line that borders the gapless phase (blue region), the Bloch band topology fails to predict the onset of interface states [phase IV] as seen in Fig. \ref{fig3}(a). In other words, the entire gapped pink zone of the phase diagram constitutes the phase space sustaining non-Hermitian non-Bloch edge states. \textcolor{black}{In addition, when increasing the length of the finite lattice, the solid line never converges with phase III, but transitions well below it [\onlinecite{suppl}].} To relate these properties to the non-Hermitian skin effect, we embark on investigating the eigenfields at the marked representative phases. \textcolor{black}{As Fig. \ref{fig3}(b) illustrates, within the conventional Hermitian phase I, all bulk states are fully extended within the entire crystal, and as usual, the in-gap  edge state quickly decays from left to right (by virtue of $Z_{A}<Z_{B}$ [\onlinecite{suppl}]) with growing lattice site number}. However, for the phases II, III and IV, the non-Hermitian nonreciprocity inhibits acoustic energy from flowing towards the positive $z$-direction, giving rise to an anomalous bulk states confinement. Alone by the shape of this skin state as rendered in Fig. \ref{fig3}(b), one can infer a violation of the translational symmetry and therefore a breaking with the common notion of Bloch-periodicity. Furthermore, the conservation of energy implies the unitarity of the transfer matrix. The magnitude of the determinant $\left| \det{(\textbf{T})} \right|$ of Eq. (\ref{T-matrix}), which is computed in Fig. \ref{fig3}(c), shows the non-unitary fingerprints of the skin effect, in that both bulk and edge states with growing attenuation, confine stronger at the left crystal-termination, resulting in a shrinking acoustic skin depth of the eigenfields in Fig. \ref{fig3}(b), from phase I to IV. \\
\begin{figure}[htbp]
	\centering
	\includegraphics[scale=1.0]{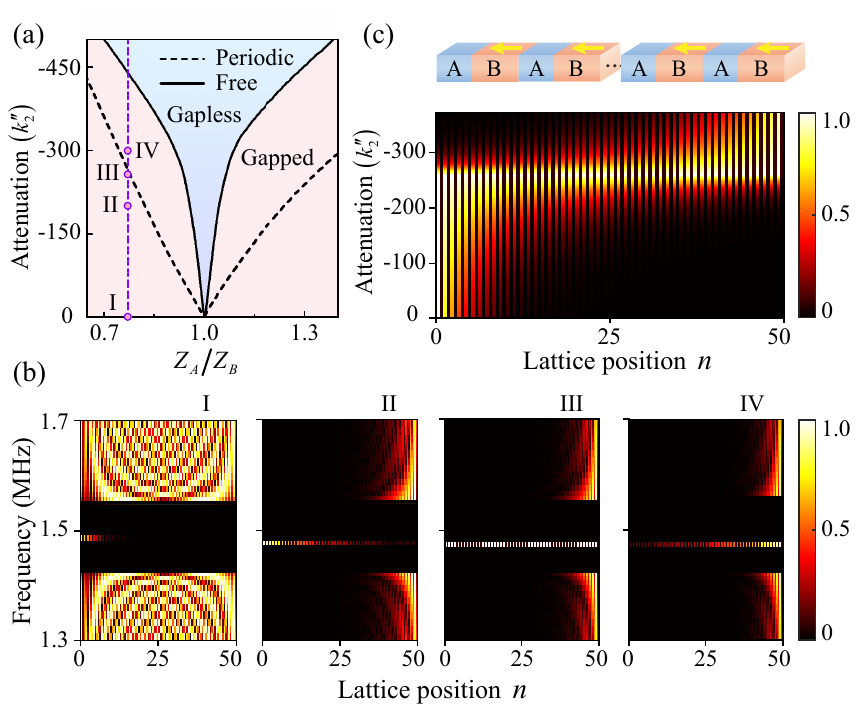}
	\caption{Non-Hermitian topology and anomalous edge states when the electric field is reversed. (a) A complete phase diagram plotted against the impedance ratio $Z_{A}/Z_{B}$ and attenuation $k''_{2}$ (in units $\text{m}^{-1}$). (b) Numerically calculated eigenfields for the selected phases as indicated in panel (a). (c) Calculated eigenfields of the in-gap edge states vs. $k''_{2}$ (dashed line in panel (a)) illustrating unusual transitions of the field confinement.}
	\label{fig4}
\end{figure}
In the next case study, we reverse the applied electric field in the piezophononic medium, i.e., $E_{0}=-v_{0}/\mu_{n}$. In such situation, the attenuation axis controlling the topological phase transition is replaced by $k''_{2}$ according to the convention discussed in Fig. \ref{fig1}. The phase diagram depicted in Fig. \ref{fig4}(a), appearing rather similar to the previous one, again demonstrates the breakdown of the Bloch band topology. In Fig. \ref{fig4}(b) however, the calculated eigenfields for the representative phases show some drastic difference. First of all, based on the reversal of $E_{0}$ bulk states confine at the right boundary. Yet, the edge states, for the specifically selected attenuation values, appear to display their very own non-Hermitian topological phase transition. As we can see in Fig. \ref{fig4}(c), the topological edge states experience an unusual \textit{near-edge}$-$\textit{bulk}$-$\textit{far-edge} transition along with an increase of $|k''_{2}|$. Particularly, what clearly stands out is the critical phase III [see Fig. \ref{fig4}(b)], at which the in-gap edge state becomes fully delocalized, while all bulk states remain localized$-$a truly non-Hermitian but anomalous skin effect.\\
\begin{figure}[htbp]
	\centering
	\includegraphics[scale=1.0]{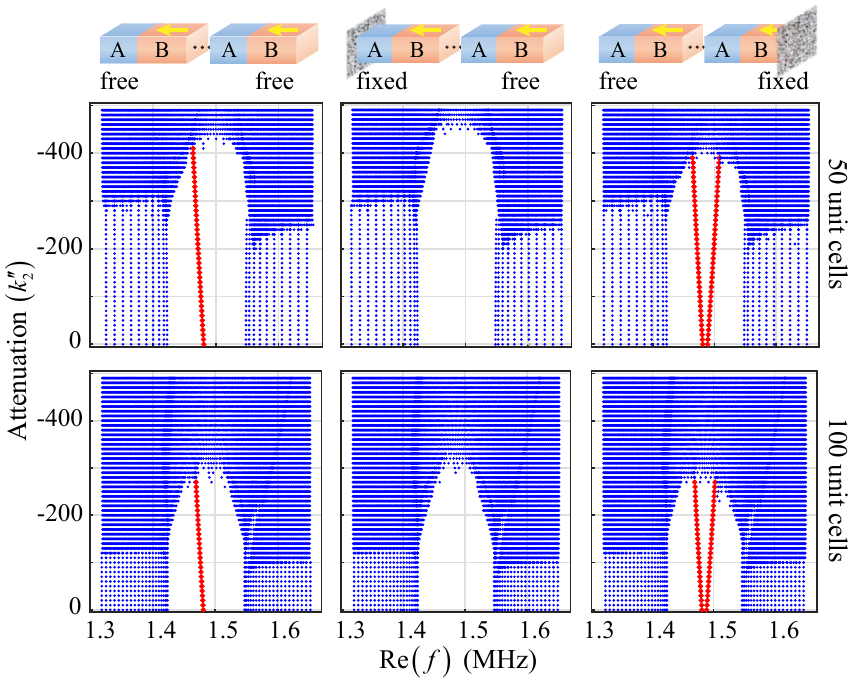}
	\caption{Numerically calculated eigenfrequency spectra subject to a variety of different boundary conditions: \textit{free-free} (left panels), \textit{fixed-free} (middle panels), and \textit{free-fixed} (right panels). Top and bottom panels correspond to finite arrays of lengths $N=50$ and $N=100$, respectively. The in-gap edge states are highlighted by red dots.}
	\label{fig5}
\end{figure}
We discussed how the non-Hermitian skin effect drastically reshapes the eigenfrequency spectra of finite systems. In this manner, we take the study one step further by examining the influence of different boundary conditions. Apart from the free case study (left panels), two additional configurations are considered, which comprise samples with the left (right) end fixed and the right (left) one free, as sketched in the middle (right) panels of Fig. \ref{fig5}. On top of that, we will also inspect the influence of the sample length, by choosing $N=50$ and $N=100$ in the top and bottom panels, respectively.  Right away, for all three problems, we see that the shorter sample requires higher levels of damping before the gap closure sets in. This is unusual, but intuitive since more attenuation is required in a shorter sample to dampen acoustic energy. The \textit{free-free} problem hosts one edge state, reminiscent to the example presented in Fig. \ref{fig2}(d). However, when the non-Hermitian piezophononic crystal is treated as a \textit{fixed-free} problem, it appears to morph into a topologically trivial system incapable to sustain edge states. In the final, \textit{free-fixed} problem, a pair of highly anomalous topological edge state materialize within the nontrivial band gap, which appear to coalesce toward the Hermitian limit ($k''_{2}=0$). These three cases constitute scenarios of non-Hermitian topology and their unusual sensitivity to the crystal terminations that have no counterpart in Hermitian systems. \\
In conclusion, by utilizing the acoustoelectric effect in layered piezophonic semiconductors, we have demonstrated the non-Hermitian skin effect and explored unparalleled topological physics for ultrasonic vibrations. Surprisingly, we show that such complex topological structure behaves utterly different compared to their Hermitian counterpart, in that Bloch bands cannot accurately predict wave properties in those finite non-Hermitian systems. Non-Hermiticity in terms of losses are usually considered a drawback in controlling sound and vibrations. Here, the tunable attenuation offers a flexible way to manipulate topologically confined edge states in finite geometries, in that losses can trap those states at either face of the sample or even in between. Beyond their fundamental significance, we foresee that the skin states and anomalous edge states presented here may stimulate investigations in ultrasonic devices for topologically robust and reflectionless signal guiding.  

\begin{acknowledgments}
J.C. acknowledges the support from the European Research Council (ERC) through the Starting Grant No. 714577 PHONOMETA and from the MINECO through a Ram\'on y Cajal grant (Grant No. RYC-2015-17156). We thank Li-Yang Zheng for helpful discussions. 
\end{acknowledgments}

\bibliography{references}

\end{document}